\documentclass[twocolumn,prl,showpacs,aps,floatfix]{revtex4}
\usepackage{graphics}
\usepackage{epsfig}
\usepackage{amsmath}

 \newcommand{\beq}{\begin{equation}}
 \newcommand{\eeq}{\end{equation}}
 \newcommand{\bea}{\begin{eqnarray}}
 \newcommand{\eea}{\end{eqnarray}}
 \newcommand{\bali}{\begin{align}}
 \newcommand{\eali}{\end{align}}

\newcommand{\bmu}{\mbox{\boldmath${\mu}$}}

\begin{document}
 
 \title{Motion-reversal in a simple microscopic swimmer }
\author{A. Alexander-Katz}
\affiliation{Laboratoire Mati\`ere Molle \& Chimie,  ESPCI, CNRS, UMR 7167, 10 rue Vauquelin, 75005 Paris, France.}

\date{\today} 

\begin{abstract}

We study the motion of a microscopic swimmer composed of a semiflexible polymer
anchored at the surface of a magnetic sphere using hydrodynamic simulations and scaling arguments. The swimmer is driven by a rotating magnetic field, and displays forward and backward motion depending on the value of the rotational frequency. In particular, the system exhibits forward thrust for frequencies below a critical frequency $\omega^*$, while above  $\omega^*$ the motion is reversed. 
\end{abstract}


\maketitle

Since the discovery of the microscopic world, scientists have tried to understand the directed motion
of the organisms that live at this scale. In particular, it has been found that many of the schemes for propulsion at low Reynolds numbers rely on beating or rotating flexible filaments periodically \cite{art:berg-phystoday,review:propulsion}. In the case of rotation, the filaments typically have helical shapes (e.g. flagellae), and move through the fluid like a cork-screw in the direction dictated by the sense of rotation. 
Further studies in this area showed theoretically 
\cite{art:goldstein-wiggins, art:manghi} and experimentlally \cite{art:mit} that simply rotating a straight semiflexible filament at its base is enough to achieve rectified motion along the axis of rotation. Within this propulsion scheme, the direction of motion turns out to be always from tail to head (i.e. the base). On the other hand, it has been recently shown by Dreyfus {\it et.al.} \cite{art:bibette}  that an actuated magnetic filament attached to a cell moves in the opposite direction, i.e. from head to tail. These results suggest that there is still much to be learned about the coupling of the hydrodynamic and elastic interactions in model microscopic swimmers, to which we refer more simply as microswimmers.  

\begin{figure}[b]
  \vspace*{-.70cm}
\begin{center}
 \includegraphics[width=80mm]{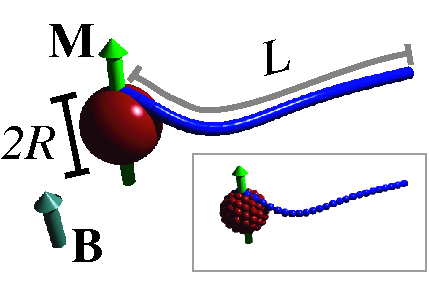}  
\end{center}
 \vspace*{-.70cm}
  \caption{The proposed microswimmer is composed of a magnetic bead (red sphere) with a semiflexible  polymer (blue cylinder) anchored at its surface. The bead carries a permanent magnetic moment 
  $\mathbf{M}$ (depicted as the green arrow), and is driven by an external magnetic field $\mathbf{B}$ (gray arrow).
Inset: Look of the swimmer after the discretization of the different components.}
  \label{fig:sketch}
  \vspace*{-.0cm}
\end{figure}

Here, we introduce a novel microswimmer that moves in both directions depending solely on the value of the driving frequency. Our swimmer is composed of a semiflexible filament (tail) that is anchored at the surface of a spherical magnetic bead (head). The swimmer is then driven by a rotatory magnetic field (see Fig. \ref{fig:sketch}). We have chosen this design because it is simple and could in principle be assembled in the laboratory. For example, one could self-assemble this system by using Streptavidin coated magnetic beads together with Taxol stabilized Biotinynlated microtubules \cite{art:dogterom}, or it can be created at the lab scale using a similar approach as that employed by Hosoi and co-workers to study low Reynolds number propulsion\cite{art:mit}. Also, our design is versatile and can be eventually used for pumping\cite{art:kim}, as well as for local mixing in microfluidic chambers, and does not have the drawback of using live non-controllable microorganisms \cite{art:brown-bacteria}. 

The origin of the forward motion displayed by our model microswimmer is due to the force that the filament produces at the anchoring point, rectified by the magnetic torque acting on the sphere. The former force is a result of the bending stiffness of the polymer, which maintains a stationary deformed shape in order to minimize the drag from the fluid. The mechanism responsible for the switching dynamics from forward to backward motion is an instability that occurs when the magnetic moment lags the field vector by a phase $\phi > \pi/2$. This occurs when the driving frequency is sufficiently large such that the maximal torque applied to the system is not enough to overcome the drag on the rotating sphere + filament. At this frequency, the magnetic dipole starts displaying large periodic oscillations along the direction perpendicular to the plane of rotation that give rise to a mix pattern of beating and rotation of the filament such that overall the system exhibits reverse motion.

We  study the motion of the proposed microswimmer by means of hydrodynamic simulations on a discretized version of our model. In particular, the magnetic bead of radius $R$ and permanent magnetic dipole moment $\mathbf{M}$ is assembled using a set of $100$ spherical particles of radius $a_s$\cite{comment:sphere}. The semiflexible filament of length $L$ and persistence length $l_p$ is also discretized into $N$ different particles of radius $a_f$ (see inset of Fig. \ref{fig:sketch}). For simplicity, we assume $a_s=a_f=a$, which does not affect the overall behavior. The polymer is anchored by one of its ends to the south (or north) pole of the sphere, and held in place by harmonic springs on the first two beads. We have considered this set-up in order to simplify the large design space, yet we note that other anchoring points or orientations may well give rise to an even richer dynamical landscape. 
 
The dynamics of the system is governed by   
\begin{equation}
\label{eq:langevin_position}
\frac{\partial}{\partial t} \mathbf{r}_i = 
 \sum_j \bmu_{ij} \cdot {\mathbf{F}_j (t)},
\end{equation} 
where $\mathbf{r}_i$ is the position of the $i$th bead, $\bmu_{ij}$ is the mobility matrix between all the particles in the system, and $\mathbf{F}_j$ is the total force acting on the $j$th bead at time $t$. The mobility matrix $\bmu_{ij}$ is taken to be the Rotne-Prager tensor.
The force on each of the particles composing the sphere consists of two terms: $\mathbf{F}=\mathbf{F}_U + \mathbf{F}_{B}$. The former corresponds to a harmonic force acting on each small bead, and can be written as $\mathbf{F}_U = -\nabla_{\mathbf{r}} U_{s}$, where $U_{s}= k  \sum (r_{ij} - d_{ij})^2$.  In the last expression, $r_{ij}$ denotes the distance between particles $i$ and $j$, and $d_{ij}$ corresponds to the "equilibrium" distance between both of these particles\cite{comment:sphere}. The spring constant is set to $k=200 k_B T/a^2$. The magnetic force $\mathbf{F}_B=\gamma \mathbf{B} \times \mathbf{M} \times \mathbf{r}_j/R$ arises from the torque on the sphere due to the magnetic field that we consider to be given by 
$\mathbf{B}=B (0,\sin (\omega t),\cos( \omega t))$, where $B$ is the magnitude of the field, and $\omega$ is the driving frequency. The numerical prefactor $\gamma$ is chosen such that the overall torque on the sphere 
is equal to $\mathbf{B} \times \mathbf{M}$.
In the case of the filament,  the force on each of the particles results from changes in the potential energy described as $U_{fil}=U_s+ U_b$, where 
$U_s$ is a harmonic potential between adjacent beads with an equilibrium distance of $2a$, and a spring constant equal to that of the springs linking the particles on the sphere. The bending potential is $U_b= \epsilon \sum_i (1-\cos \theta_i)$, where $\epsilon$ measures the stiffness of the filament, and $\theta_i$ is the angle between adjacent bond vectors $\mathbf{u}_i =\mathbf{r}_{i+1}-\mathbf{r}_i   $.  The persistence length follows as $l_p = a \epsilon /k_B T$.  
Finally, we re-scale all the lengths by $R$, the forces by $k_B T/a$, and the time by the characteristic diffusion time of the sphere $\tau_R=R^2/ (\mu_R k_BT)$, where $\mu_R=1/(6 \pi \eta R)$ is the Stokes mobility of the magnetic bead. The dimensionless parameters that govern the dynamics  of the microswimmer are then the rescaled length of the filament $\tilde{L}=L/R$, the dimensionless persistence length $\tilde{l}_p = l_p /R$, the maximum magnetic torque (in units of $k_BT$) $\tilde{\alpha}_{max} = BM/k_BT$, and the dimensionless driving frequency $\tilde{\omega}=\omega \tau_R$.
 \begin{figure}[t]
\begin{center}
   \includegraphics[width=80mm]{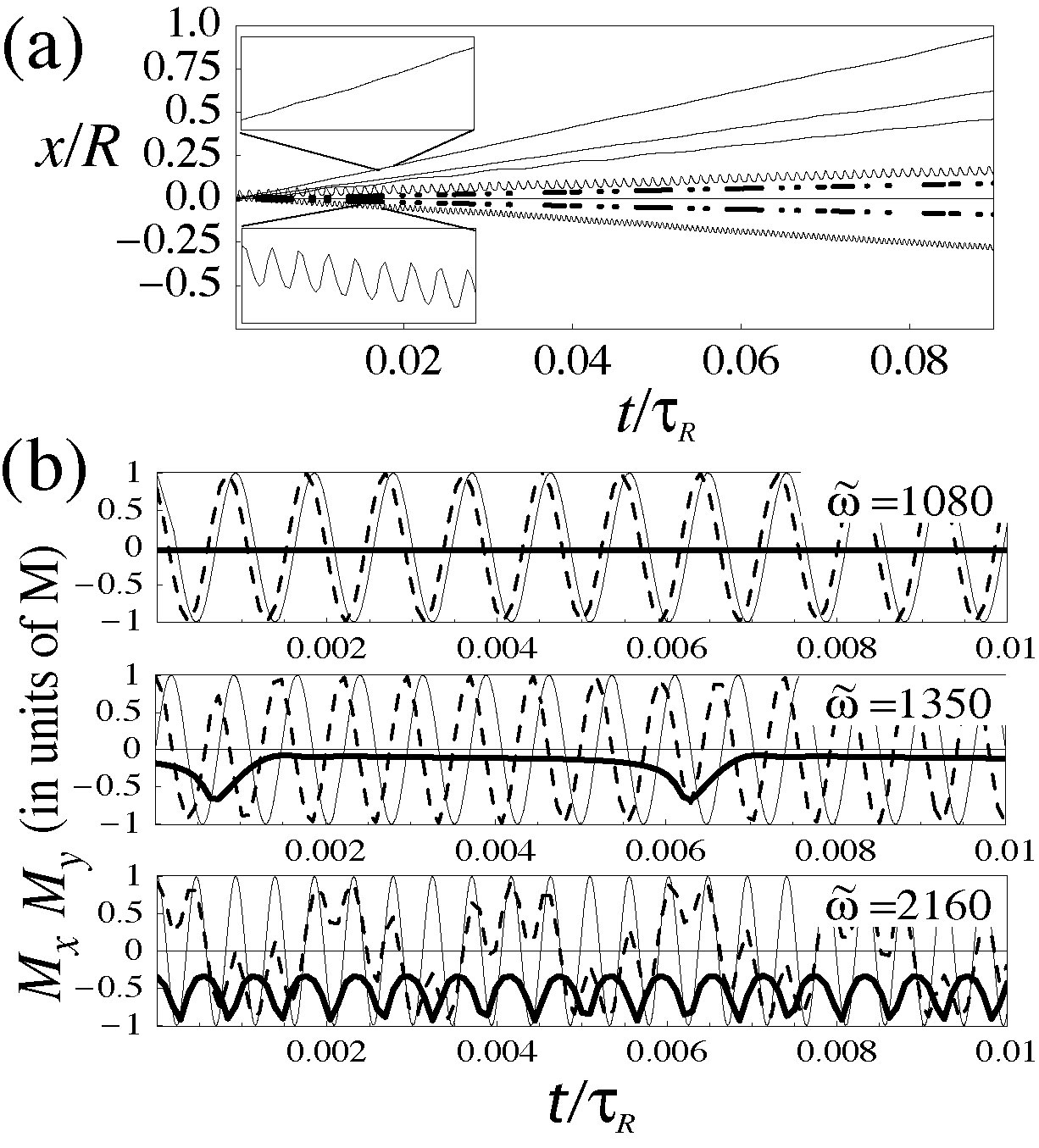}  
\end{center}
 \vspace*{-.50cm}
  \caption{(a) The distance (in units of the sphere radius $R$) traveled by the microswimmer in the $x$ direction as a function of the rescaled time $t/ \tau_R$. The magnetic field and moment are held fixed such that the maximum torque $\tilde{\alpha}_{max} \approx 9500$. The filament length is $\tilde{L} \approx 8.3$, and the persistence length is given by $\tilde{l}_p \approx 160$. The different solid curves correspond to $\tilde{\omega}  =1080,108,54,1620,2160 $ (from top to bottom). The dot-dashed curves correspond to the diffusive motion of the magnetic bead. Insets: amplification of the trajectories for the cases  $\tilde{\omega}  = 1080 $(upper inset), and $2160$ (lower inset). (b) Time sequences of two components of the permanent dipole moment $\mathbf{M} $ of the sphere at three driving frequencies (indicated in the upper right corner of each graph). The thick solid curves correspond to the $x$ component $M_x$ and the dashed lines to the $y$ component $M_y$. The thin solid line denotes the value of $B_y(t)/B$. All the sequences shown here have the same parameters as in part (a). }
  \label{fig:trajectories}
  \vspace*{-.50cm}
\end{figure}

In Fig. \ref{fig:trajectories} (a) we show five different trajectories in the $x$ direction for a set-up  characterized by  $\tilde{\alpha}_{max} \approx 9500$, $\tilde{L} \approx 8.3$, and $\tilde{l}_{p} \approx 160$. As noted before, the swimmers propels forward (positive slope) for values of $\omega < \omega^*$, which for this particular case corresponds to $\omega^* \sim 1350$ (see below).  For frequencies above $\omega ^*$ the velocity diminishes considerably and starts displaying reverse motion (negative slope).   Notice that the trajectories are smooth for $\omega  <\omega^*$, but above this frequency the system displays periodic forward and backward motion, that results in a net negative speed if $\omega$ is high enough (see lower inset). The origin of this behavior is due to the fact that the sphere not only rotates, but also "rocks" on the filament, and this produces a beat-like pattern. Actually, we define the value of $\omega^*$ as the point at which the swimmer starts displaying beating motion, which is directly related to the appearance of "strokes" of the permanent magnetic dipole along the $x$ axis. To illustrate this events, we present time sequences of $\mathbf{M}$ at frequencies below, at, and above the transition in Fig. \ref{fig:trajectories} (b). From these sequences we first notice that the $y$ component $M_y$ of the dipole moment follows the 
driving field up to the transition point (upper and middle sequence). At this particular value of $\tilde{\omega}= \tilde{\omega}^*$, the $x$ component of the permanent magnetic moment $M_x$ starts displaying large strokes at intervals quite larger than the period of oscillation of the field (see middle sequence). This behavior is retained for frequencies larger than $\omega^*$, except that the strokes become more frequent (lower sequence). Also, note that $M_y$ seems to be a periodic function with a period larger than $1/\tilde{\omega}$ modulated by the field. The structure of $M_y$ is nevertheless complicated and cannot be written in simple terms.   
However, the onset of the complex dynamics is found to correlate well with the appearance of large amplitude periodic strokes in $M_x$.

\begin{figure}[t]
\begin{center}
   \includegraphics[width=80mm]{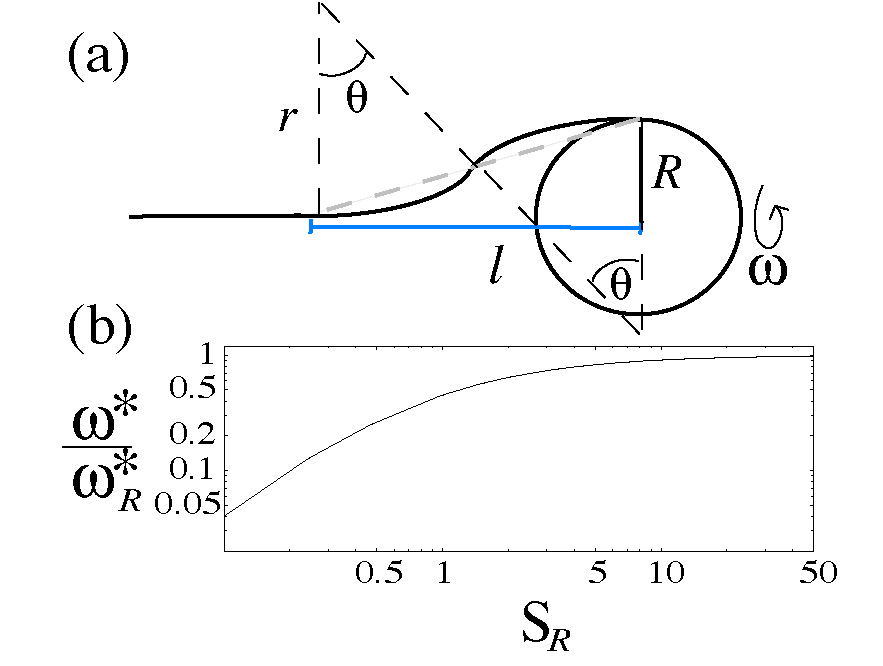}  
\end{center}
 \vspace*{-.50cm}
  \caption{(a) Model of the stationary state of the microswimmer. The shape of the deformed part of the filament is described by two pieces of a circle of radius $r$. The overall deformation length is $l$, and $R$ is the radius of the sphere. (b) The dependence of the rescaled critical frequency $\omega^*/\omega_R^*$ as a function of the characteristic sperm number $S_R \sim R/a (\omega_R^* \tau_{\perp} a/l_p)^{1/4}$. The critical frequency in the absence of the filament is defined as  $\omega_R^* = BM/\xi_r$.}
  \label{fig:model}
  \vspace*{-.50cm}
\end{figure}

To rationalize this behavior, and in particular estimate the critical frequency  ${\omega}^*$, we introduce a simple model for the microswimmer (see Fig. \ref{fig:model} (a)). The most important assumption within our rationalization is that the dynamics of the system is stationary, and thus the deformation of the filament shape is constant in time. By minimizing an "energy" functional 
$\mathcal{F}=  \mathcal{E} + \mathcal{W}$ that corresponds to the sum of the deformation energy 
$\mathcal{E}$ plus the work done on the system per unit cycle $\mathcal{W}$ we calculate the deformation length $l$. Finally,  by comparing the maximum torque with the dissipation of the sphere+filament system we obtain the critical frequency $\omega^*$. The energy of deformation within our model is given by $\mathcal{E} \sim l_p R^2 k_B T /l^3$, where we have considered that the deformed piece of the filament can be described by two arcs of a circle of radius $r$ each spanning an angle $\theta$. This simple result captures to leading order  in $l$ the deformation energy. The work done on the filament  per cycle is given (to leading order in $l$) by  $\mathcal{W} \sim \omega \tau_{\perp} k_B T R^2  l/a^3$, where $\tau_{\perp}=\xi_{\perp} a^2 /k_B T$, and $\xi_{\perp}$ is the perpendicular drag constant of a filament segment of length $a$.
Minimizing $\mathcal{F}$ yields the deformation length 
$l \sim a (l_p /(a \omega \tau_{\perp}))^{1/4}$. The critical frequency is then simply calculated by setting the maximum torque $BM$ equal to the dissipation of the rotating sphere+filament $\xi_R \omega + \mathcal{W}$. This results in the  following equation for $\omega^*$
\begin{equation}
\frac{\omega^*}{\omega_R^*}+S_R^{-1}\left( \frac{\omega^*}{\omega_R}\right)^{3/4}=1,
\label{eq:frequency} 
\end{equation}
where  $\omega_R^*=BM/\xi_R$ is the critical frequency for a single magnetic bead with no grafted filament, and $\xi_R$ is the rotational drag coefficient of the sphere. Note that this equation depends only on a single dimensionless parameter $S_R \sim R/a (\omega_R^* \tau_{\perp} a/l_p)^{1/4}$ that corresponds to a characteristic "sperm number". This parameter can be conveniently written  also as  $S_R \sim (\tilde{\alpha}_{max} /\tilde{l_p})^{1/4}$. The behavior of  $\omega^*/\omega_s^*$ as a function of $S_R$ is shown in Fig. \ref{fig:model} (b). As expected, for large values of $S_R$ (low rigidity or large magnetic fields) $\omega^*/\omega_R^*$ is essentially 1, while for values of $S_R < 1$ the 
critical frequency starts deviating considerably from  $\omega_R^*$ and can be easily shown to scale as $\omega^* \sim \omega^*_R S_R^{4/3} ~ \sim \omega^*_R (\tilde{\alpha}_{max} /\tilde{l_p})^{1/3}$ . Nevertheless, it should be noted that the dependence on the physical parameters is very weak. Also, it is important to point out that we have neglected all the numerical prefactors in our derivation since our theory only depends on a single parameter whose functional form we obtained. The prefactor for $S_R$ will in general be system dependent, and can be calculated by computing $\omega^*/\omega_R^*$ from the simulations (or experiments)  at a fixed value of $\tilde{\alpha}_{max}$, as is done in what follows.    

\begin{figure}[t]
\begin{center}
   \includegraphics[width=80mm]{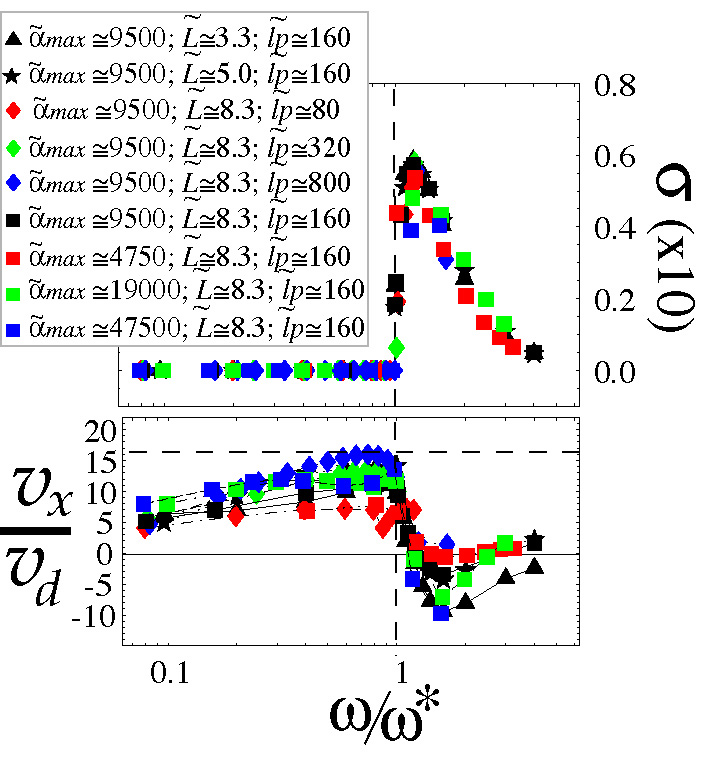}  
\end{center}
 \vspace*{-.50cm}
  \caption{(a) The variance $\sigma = \langle(M_x/M)^2 \rangle - \langle(M_x/M)\rangle^2 $,  and (b) the rescaled velocity $v_x / v_d $ (see text for the definition of $v_d$) as a function of the rescaled driving frequency  $ \omega/\omega^*$. The different parameters used in each curve are displayed in the legend. The horizontal dashed line corresponds to the maximal achieved velocity $v_{max}^{fwd} \sim 16 v_d$. The vertical dashed line corresponds to the critical frequency at which the system becomes unstable (see text for details).}
  \label{fig:velocity}
  \vspace*{-.30cm}
\end{figure}  

Finally, in Fig. \ref{fig:velocity} (a) we present the variance of the $x$ component of the magnetic dipole $\sigma = \langle(M_x/M)^2 \rangle - \langle(M_x/M)\rangle^2 $  as a function of $\omega/\omega^*$. The  variance $\sigma$ is very sensitive to the appearance of oscillations in $M_x$, and could be measured experimentally using induction techniques. The critical frequency $\omega^*$ for the different parameters employed here is calculated from Eq. \ref{eq:frequency} using  
$S_R = 4.8 (\tilde{\alpha}_{max} /\tilde{l_p})^{1/4} $. The prefactor is obtained from the data at $\tilde{\alpha}_{max} \approx 9500$, $\tilde{L} \approx 8.3$, and $\tilde{l}_p \approx 160$.  It can be clearly observed from this graph that the magnetic moment suddenly starts "oscillating" in the $x$ direction with a large amplitude since $\sigma$ jumps sharply at the critical frequency. For larger frequencies, the amplitude of these oscillations becomes smaller, and hence the decrease in $\sigma$. Interestingly, all the curves collapse when plotted in this way, implying that the dynamics of the oscillations only depends on $\omega/\omega_*$.
In Fig. \ref{fig:velocity} (b) we plot the corresponding velocites in the $x$ direction $v_x/v_d$ 
for the same parameters as in part (a). The velocity  $v_d = R/\tau_R$ corresponds to the diffusive speed of the magnetic bead. As can be observed from this graph, the velocity grows as a function of the frequency for $\omega < \omega^*$. Near (but below) $\omega^*$ it essentially plateaus, and finally above the critical frequency it undergoes an abrupt decline in its value until negative velocities are attained.  Interestingly, as the frequency is tuned to even higher values the swimmer starts moving forward again. The maximum forward velocity found for the parameters employed here is $v_{max}^{fwd} \sim 16 v_d$, for the case $\tilde{\alpha}_{max} \approx 9500$, $\tilde{L} = 8.3$, and $ \tilde{l}_p = 800$. In a real set-up, these parameters would be equivalent to having a microtubule of length $L \sim 10 \mu$ grafted on a $1 \mu$ radius bead, and swimming at a speed of  $ v \sim 8 \mu/s$. In the reverse motion we found $v_{max}^{rev} \sim 10 v_d$ for the following parameters: $\tilde{\alpha}_{max} \approx 47500$, $\tilde{L} = 8.3$, and $ \tilde{l}_p = 160$. These results suggest that the fastest forward swimmers will in general not move accordingly in the opposite direction above the transition frequency. Also, we note that the maximum velocities will be affected by the cross section of the filament, which in general will be quite smaller than the one considered here. Reducing the cross-section reduces the drag, and thus increases the final velocity.  In particular, we note the drag on our model swimmer comes mostly from the tail since it contributes $\sim1/(1 +  \tilde{L}^{-1})$ of the overall friction coefficient (neglecting hydrodynamic interactions). Reducing this quantity by a factor of 5 should yield velocities that are of the order of 100x larger than its typical diffusion motion. Thus, it could be foreseable to use this type of microswimmer for perfoming tasks at the micron scale with high precision. 

Our results also highlight the difference for externally actuated microswimmers between constant driving force (or torque) and constant driving frequency at fixed field strength, since they lead to different 
qualitative behavior. In the particular case studied here, the former "ensemble" would have not led to any instability, while as shown here, the latter "ensemble" does exhibit a non-intuitive motion-reversal transition. This point is relevant because in most situations one deals with constant driving frequencies, and not with constant torques. Apart from this, our results also show that increasing the value of the field does not lead to large changes in the maximum velocity that the microswimmer can attain.    
 Another important point is due, and it concerns the anchoring of the filament. In all this work we have assumed that the semiflexible polymer can relax the torsion along its contour, or that the anchoring is
 "slippy". Although this type of attachment can be constructed, a more realistic scenario is a grafted filament completely fixed at the base. In this case, and assuming that the polymer cannot relax the stresses by rotation around its backbone, the torsion will not be zero and can lead to the so-called twirling-whirling instability  \cite{art:goldstein-powers}, and many other interesting phenomena such as plectonemes \cite{art:plectonemes}. Nonetheless, one can estimate that this transition will occur at $\omega_{twi} \sim 5 \omega^*$ for our fastest swimmer. Hence, our results should be clearly observed in the laboratory, even if the filament is not allowed to rotate freely around its backbone.        
 
In summary, we have introduced a simple, yet novel design for an externally driven microswimmer that could be self-assembled in the lab. The swimmer displays forward or  backward motion depending on the value of the frequency of the actuating magnetic field. In particular it undergoes a transition at a critical driving frequency $\omega^*$ above which it starts displaying stroke movements in the direction perpendicular to the plane of the rotating field. The maximum velocities achieved are well above the typical diffusion velocity, and thus the swimmer can be controlled to up to very small scales.
The design proposed here is versatile and could be easily employed to study swarming behavior or hydrodynamic synchronization. Also, arrays of these swimmers could be used as ciliae-like biomimetic pumps in microfluidic applications.

The author would like to thank Roland R. Netz and Hirofumi Wada for illuminating discussions. Financial support from the NSF is acknowledged.


\begin{thebibliography}{10}
\bibitem{comment:sphere}
The shell is constructed by allowing $100$ repulsive particles of radius $a$ to diffuse freely on the surface of the sphere, while slowly turning the temperature down. The final structure at 
$T =0$ is fully "cross linked" by harmonic springs between all the beads.
\end{thebibliography}
 \end{document}